\documentclass[letterpaper,twocolumn,10pt]{article}
\usepackage{usenix}

\usepackage{tikz}
\usepackage{amsmath}
\usepackage{booktabs}
\usepackage{amssymb}
\usepackage{bbm}
\usepackage{filecontents}

\begin{document}

\date{}

\title{\Large \bf MixedTruth: Component-Level Detection of \\Synthetic Speech and Synthetic Background in Mixed Audio}

\author{
{\rm Xindong Zheng}\\
Vanderbilt University
\and
{\rm Jiashu Zhu}\\
Vanderbilt University
\and
{\rm Ke Li}\\
Vanderbilt University
\and
{\rm Dan Lin}\\
Vanderbilt University
} 

\maketitle

\begin{abstract}
Modern generative audio can fabricate not only voices but also the acoustic environments that provide evidence of when and where speech occurred. This capability undermines a longstanding forensic assumption that background sound provides reliable evidence of a recording's provenance. Yet existing detectors produce only a single real-or-synthetic verdict, leaving analysts unable to determine whether the voice, the environment, or both were manipulated. In this paper, we propose a component-level forensic detector, namely TruMix, which classifies recordings into eight categories spanning the presence and authenticity of speech and background. TruMix has a unique multi-path architecture that jointly analyzes speech-oriented and background-oriented spectral evidence, contextual information from the mixed waveform, and low-level synthesis artifacts. This design avoids the artifact propagation inherent in conventional separate-then-detect pipelines while enabling generalization across independent speech and environmental-sound generators. We also curated MixedTruth, a 4,917-hour benchmark covering all eight categories under diverse mixing conditions. Extensive experiments show that TruMix achieves 98.51\% AUC for component-level attribution, significantly outperforms ten state-of-the-art synthetic-speech detectors in conventional binary classification, and exhibits strong robustness across four attack tiers.
\end{abstract}

\section{Introduction}
For most of the past decade, an ``audio deepfake'' has effectively meant a
``synthetic human voice''.  Threat models, benchmark datasets, and countermeasures have
largely focused on  deciding whether that utterance was produced by a person or
a machine~\cite{yamagishi2021asvspoof, wang2024asvspoof, frank2021wavefake}. This premise no longer
captures the capabilities of modern generative audio. Recent text-to-audio
systems can synthesize environmental sounds with striking realism, such as rain
against a window, traffic on a street, a crowd inside a stadium or at an event scene, or
birdsong in a park~\cite{kreuk2022audiogen, liu2023audioldm, evans2024fast}. An adversary can
therefore fabricate not only what a speaker appears to have said, but also
the entire acoustic world in which the statement appears to have been made.

This expansion fundamentally changes the security problem because background
sound is not merely interference, but also evidence that provides  \emph{when} and \emph{where} the speech  occurred. For instance, an impostor can embed synthetic speech in a plausible environment to make an
otherwise artificial recording appear more trustworthy. A cloned voice in implausible
silence may invite scrutiny, but  the same voice mixed with cafe chatter, passing
traffic, and appropriate room reverberation can make the audio appear to be from a normal phone call. On the other hand, an adversary can place genuine speech into a synthetic
acoustic environment, making a real person appear to have spoken at a
protest, campaign event, combat zone, or other location they never visited.

Existing countermeasures are inadequate for this expanded threat. Most synthetic speech
detectors are trained and evaluated on clean or relatively
close-talking speech, and, as our experiments show, their performance 
degrades sharply when speech is mixed with environmental sound. Even when an existing
 detector occasionally identifies synthetic content, it produces only
a single verdict for the entire recording. Such a verdict cannot tell an
investigator whether the speaker was impersonated, the acoustic scene was
fabricated, or both components were generated. To
our knowledge, no prior work provides such component-level authenticity
analysis for composite speech-background audio.

To address these new challenges, we propose {TruMix}, a detector that
jointly determines the presence and authenticity of speech and background
within an input recording and produces one of the following eight component-level
verdicts: (i) real speech only (RS); (ii) real speech in real background (RSRB); (iii)  real speech with synthetic background (RSSB); (iv) synthetic speech inserted to real background (SSRB); (v)~synthetic speech inserted to synthetic background (SSSB);  (vi) synthetic speech only (SS); (vii) real background only (RB); and (viii) synthetic background only (SB).  Component-level attribution is substantially more challenging than simply separating speech from background audio and applying existing detectors to each component independently. Such a naive pipeline propagates and often amplifies source-separation artifacts, therefore degrading downstream detection accuracy. More fundamentally, the speech and background components may be produced by different generators, which requires the detector to generalize simultaneously along two independent axes of synthesis rather than a single one.

TruMix has a unique multi-path architecture that examines the same recording from four complementary forensic perspectives
and fuses their evidence for joint component-level attribution. Rather than
simply separating speech from background and applying an independent detector
to each estimated source, {TruMix} simultaneously analyzes
(i) speech-oriented spectral evidence, (ii) background-oriented spectral
evidence, (iii) the contextual information preserved in the original mixed
waveform, and (iv) low-level waveform artifacts indicative of audio synthesis.
These paths capture different and potentially complementary manifestations of
manipulation, including spectral irregularities associated with a particular component,
inconsistencies between speech and its acoustic surroundings, global acoustic
context, and subtle generation traces that may be obscured in higher-level
representations. Their fusion enables {TruMix} to determine not only
whether a recording contains synthetic content, but also which component was
manipulated.


In summary, we make the following contributions: 
\begin{itemize}
\item We formulate \emph{component-level authenticity analysis} for
composite speech-background audio and introduce an eight-way classification
covering all valid combinations of speech and background presence and
authenticity.

\item We design  {TruMix}, a multi-path forensic architecture that
fuses four complementary views of an audio recording. Unlike a conventional
separate-then-detect pipeline, {TruMix} jointly reasons across source-specific, contextual, and forensic
evidence.

\item We curate MixedTruth, a new dataset comprising approximately 4,917 hours of speech–background mixtures generated at varying speech-to-background ratios and annotated according to our eight-category taxonomy. MixedTruth enables controlled evaluation of component-level presence and authenticity detection across diverse mixing conditions.

\item  We conduct extensive experiments that comprehensively evaluate {TruMix} across mixed and
single-component recordings and under various attack scenarios. Our results demonstrate both the
inadequacy of 10 state-of-the-art synthetic voice  detectors, and the advanced detection capability of the proposed  {TruMix}, e.g., 98.513\% macro one-vs-rest
AUC.
\end{itemize}

The remainder of the paper is organized as follows. Section \ref{sec:related} reviews existing synthetic voice detectors.  Section \ref{sec:threat} describes the threat model. Section \ref{sec:trumix} presents our proposed TruMix. Section \ref{sec:exp} reports experimental results. Finally, Section \ref{sec:con} concludes the paper.

\begin{table*}[!t]
	\centering
	\caption{Architecture and model complexity of state-of-the-art synthetic
		speech detector. }
	\label{tab:detector_architecture_complexity}
	\footnotesize
	\setlength{\tabcolsep}{4pt}
	\renewcommand{\arraystretch}{1.12}
	\begin{tabular}{lp{8.2cm}rr}
		\toprule
		\textbf{Detector (Year)}
		& \textbf{Main Architecture}
		& \textbf{Parameters}
		& \textbf{FP32 Size} \\
		\midrule

		TrueEar~\cite{baird2025trueear} (2025)
		& SincConv + temporal encoder + VLAD
		& 0.048M & 0.192 MB \\

		AASIST-L~\cite{jung2022aasist} (2022)
		& lightweight raw-waveform encoder + spectro-temporal graph attention
		& 0.085M & 0.340 MB \\

		AASIST~\cite{jung2022aasist} (2022)
		& Raw-waveform encoder + spectro-temporal graph attention
		& 0.298M & 1.192 MB \\

		RawNet2~\cite{tak2021end} (2021)
		& SincConv + residual blocks + GRU
		& 17.621M & 70.484 MB \\

		ALDEN~\cite{xu2025alden} (2025)
		& XLS-R (24 blocks) + dual-level disentanglement + meta-learning 
		& 316.274M & 1.265 GB \\

		Trident of Poseidon~\cite{doan2024trident} (2024)
		& XLS-R (24 blocks) +  VIB head
		& 317.572M & 1.270 GB \\

		AASIST2~\cite{zhang2024improving} (2024)
		& XLS-R (24 blocks) + Res2Net-enhanced AASIST
		& 317.838M & 1.271 GB \\

		Nes2Net-X~\cite{liu2025nes2net} (2025)
		& XLS-R (24 blocks) + nested Res2Net
		& 317.903M & 1.272 GB \\

		AASIST3~\cite{borodin2024aasist3} (2024)
		& XLS-R (24 blocks) +  KAN-enhanced AASIST
		& 321.749M & 1.287 GB \\

		XLS-R+SLS~\cite{zhang2024audio} (2024)
		& XLS-R (24 blocks) +  sensitive-layer selection
  + utterance-level classifier
		& 340.790M & 1.363 GB \\

		\midrule
		\textbf{TruMix (Ours)}
		& Conv separator
		+ dual-stream 4-block spectral Transformers
		+ 12-block contextual waveform Transformer + waveform-artifact encoder
		& 98.778M & 395.112 MB \\

		\bottomrule
	\end{tabular}
\end{table*}

	\section{Related Works} \label{sec:related}


RawNet2~\cite{tak2021end} was among the first end-to-end anti-spoofing models to operate on raw waveforms. It combines a sinc-filter front end, residual blocks, and a GRU for utterance-level classification. AASIST~\cite{jung2022aasist} subsequently introduced graph-attention modules that jointly model spectral and temporal artifacts, which improves the performance. 

Recent detectors increasingly rely on large self-supervised speech encoders. AASIST2~\cite{zhang2024improving} feeds XLS-R features into an AASIST back end augmented with multi-scale Res2Net blocks. Trident of Poseidon~\cite{doan2024trident} combines XLS-R with a variational information bottleneck and generalization-oriented training using augmentation and re-synthesized speech. XLS-R+SLS~\cite{zhang2024audio} learns to select informative XLS-R layers, while AASIST3~\cite{borodin2024aasist3} augments AASIST with Kolmogorov-Arnold Network modules and additional regularization. Nes2Net-X~\cite{liu2025nes2net} directly processes high-dimensional SSL features using nested Res2Net modules, and ALDEN~\cite{xu2025alden} improves cross-vocoder generalization through feature disentanglement and meta-learning. Although effective, these systems contain approximately 318--341M parameters and produce only binary utterance-level decisions.

TrueEar~\cite{baird2025trueear} represents the latest lightweight alternative. It combines parameterized sinc filters, temporal encoding, and VLAD-style aggregation using only 0.048M parameters. However, as shown in our experiments, this reduction in model size comes at the cost of detection accuracy. 

Existing detectors are commonly evaluated on ASVspoof \cite{wang2024asvspoof,yamagishi2021asvspoof}, WaveFake \cite{frank2021wavefake}, Fake-or-Real\cite{reimao2019dataset}, ADD \cite{yi2022add,yi2023add}, and In-the-Wild \cite{muller2022does}. VoiceWukong~\cite{yan2025voicewukong} extends these benchmarks with English and Chinese speech from 34 commercial and open-source generators and 38 post-processing variants. Its evaluation shows that existing detectors generalize poorly to current generators and realistic manipulations. However, their labels specify only whether an entire utterance is real or synthetic. Thus, for component-level detection, we introduce the new benchmark MixedTruth. 

Unlike existing binary detectors, TruMix is the first to perform fine-grained component-level synthetic-speech detection. Its component-aware predictions provide more actionable forensic evidence than a single binary decision. Moreover, with 98.778M parameters, TruMix is substantially smaller than the evaluated XLS-R-based detectors.

Table \ref{tab:detector_architecture_complexity} summarizes these state-of-the-art synthetic speech detectors which are evaluated in our experiments. Parameter counts shown in the table correspond to the implementations used in our experiments and include pretrained backbones. Model storage	is estimated using FP32 parameters (four bytes per parameter), excluding
		optimizer states and framework metadata. Baselines are ordered by ascending parameter count, with TruMix shown last.

\section{Threat Model} \label{sec:threat}

We consider an adversary who seeks to create a deceptive audio recording by
manipulating the authenticity of its speech component, its background component,
or both. The adversary pursues one or both of the following objectives:
\begin{itemize}
    \item {\bf Identity Falsification}: making synthetic speech appear authentic by embedding it in a plausible acoustic scene, whether that scene is genuinely recorded or fabricated. 
    
    \item {\bf Context Falsification}: placing genuine speech in a synthetic acoustic scene that misrepresents when and where the utterance occurred.
\end{itemize}

The adversary may obtain genuine speech and may	generate synthetic speech using text-to-speech or voice-cloning systems and 	synthetic background audio using text-to-audio systems. We study the following four attacker-knowledge and access regimes. These regimes represent progressively greater access to the deployed detector. 

\vspace{5pt}\noindent$(i)$ \textbf{Model-agnostic manipulation.}
    The adversary has neither knowledge of the detection model nor multiple query budget to it. Instead, the adversary applies plausible production and
    post-processing operations that may suppress synthesis artifacts. For instance, the adversary can mix speech and background components using arbitrary relative gains by independently adjusting each component's volume to control its contribution to the final signal.  The adversary may also apply common post-processing operations, such as lossy compression using formats or codecs which discard portions of the signal and may remove subtle forensic traces, or add reverberation to introduce simulated room echoes and acoustic reflections.

\vspace{5pt}\noindent$(ii)$ \textbf{Decision-based black-box attack.}
    The adversary can repeatedly submit an audio file and observe only the
    final discrete verdict. Note that confidence scores and model parameters are not
    available. The adversary  then searches for a    nearby input to evade the detection.  
    

\vspace{5pt}\noindent$(iii)$ \textbf{Gray-box transfer attack.}
    The adversary does not know the parameters of the deployed  detection model and does not use its outputs while constructing    the attack. The adversary trains its own surrogate model and
    generates targeted adversarial examples against the target detector. 
    

\vspace{5pt}\noindent$(iv)$ \textbf{White-box attack.}
  The adversary has complete knowledge of the target detector, including its parameters. It iteratively optimizes the input, while keeping the model parameters fixed, until the detector outputs the desired target class. This setting represents the strongest attacker considered.

The defender observes only the final waveform. Specifically, the proposed {TruMix} does not require access to the original clean speech or background, and does not have any prior knowledge of the claimed speaker, location, or event. Our objective is to determine whether the present components in the received waveform, i.e., speech and/or background are  genuine or synthetic.

{TruMix} performs signal-level synthetic-content attribution. It does not establish the semantic truth of a recording or prove that it was captured at a claimed time or location. Consequently, deceptive uses of entirely genuine audio, such as relabeling or combining genuine
recordings without introducing detectable synthetic content, are outside our
scope.

\begin{figure*}[!t]
    \centering
    \includegraphics[width=.97\textwidth]{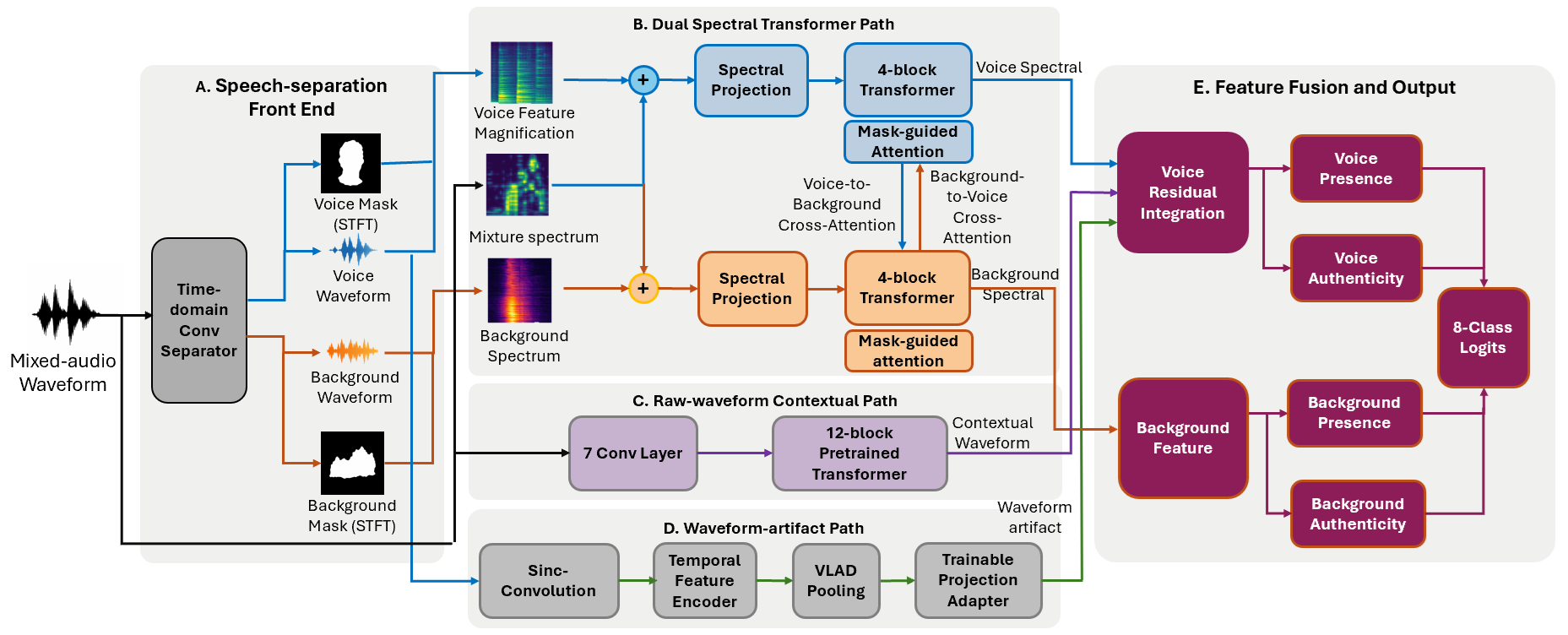}
    \vspace{-8pt}
    \caption{TruMix architecture.}
    \label{fig:trumix}
\end{figure*}

\section{Our Proposed TruMix} \label{sec:trumix}



Figure \ref{fig:trumix} illustrates an overview of TruMix architecture. The core of TruMix is a dual spectral Transformer that performs
	component-aware analysis without treating speech and background as independent
	signals. A frozen time-domain separator estimates speech and background
	waveforms and their corresponding masks. {TruMix} uses these outputs as
	attention guidance rather than accepting them as definitive decompositions of
	the recording. The speech path combines the mixture spectrum with mask-derived
	speech evidence to magnify time-frequency regions likely to contain voice,
	while the background path focuses on the estimated nonspeech spectrum.
	Mask-guided attention further directs each path toward evidence relevant to its
	target component.

The other two paths preserve evidence that may be lost or distorted during
	source separation. A pretrained raw-waveform Transformer analyzes the original
	mixture to extract global acoustic and semantic context, providing a holistic
	view of the recording before it is decomposed. In parallel, a dedicated
	waveform-artifact path employs parameterized sinc filtering, temporal feature
	encoding, and learnable aggregation to capture fine-grained synthesis traces.
	This path complements the higher-level contextual representation by retaining
	low-level forensic evidence that may be attenuated by spectral projection,
	source separation, or pretrained representation learning.
	
	Finally, TruMix fuses the speech-spectral, background-spectral,
	contextual-waveform, and waveform-artifact representations and reasons over 	their combined evidence using a factorized output design.

\subsection{Mixture-Preserving Dual Spectral Path}
\label{sec:dual_stream}

Source separation exposes useful structure but is not a reliable forensic
decision boundary. Compression, reverberation, spectral overlap, and
adversarial post-processing can suppress manipulation artifacts or displace
them across estimated sources. A detector that analyzes only separated
waveforms therefore inherits irreversible separation errors. We instead
follow three design principles: (1) preserve the original mixture as a
forensic anchor, (2) treat separated sources as soft evidence rather than
ground truth, and (3) freeze the separator to prevent detection labels from
inducing task-specific decompositions.

Given waveform \(x\), a frozen Conv-TasNet separator
\(\mathcal{S}\)~\cite{luo2019conv}, pretrained on source-supervised
ASVspoof2019-LA mixtures, estimates voice and background components:
\begin{equation}
    (\hat{x}_v,\hat{x}_b)=\mathcal{S}(x).
\end{equation}
We compute their spectra \(X,\hat{X}_v,\hat{X}_b\) using a 512-point STFT
with a 128-sample hop, producing \(F=257\) frequency bins. Because
\(\mathcal{S}\) does not expose masks, we derive soft ratio masks:
\begin{equation}
    M_c =
    \frac{|\hat{X}_c|}
    {|\hat{X}_v|+|\hat{X}_b|+\epsilon},
    \qquad c\in\{v,b\},
\end{equation}
where \(\epsilon=10^{-8}\). We represent a spectrum \(Z\) using its
per-recording, mean--variance-normalized log magnitude:
\begin{equation}
    L(Z)=
    \operatorname{Norm}\!\left(\log(1+|Z|)\right).
\end{equation}

Separation may attenuate weak forensic traces. We therefore use residual
convolutional magnifiers to refine only the voice evidence:
\begin{align}
    \widetilde{L}_v
    &=
    L(\hat{X}_v)
    +\tanh(g_X)\mathcal{R}_X\!\left(L(\hat{X}_v)\right),
    \\
    \widetilde{M}_v
    &=
    \sigma\!\left(
        M_v+\tanh(g_M)\mathcal{R}_M(M_v)
    \right),
\end{align}
where \(\mathcal{R}_X\) and \(\mathcal{R}_M\) are convolutional
encoder--decoders, and \(g_X\) and \(g_M\) are learned gates. The bounded
gates make these refinements conservative as the modules act as perturbations
of the separator outputs rather than unconstrained replacements.

The voice and background streams receive
\begin{align}
    F_v &= [L(X);\,\widetilde{L}_v;\,\widetilde{M}_v],
    \\
    F_b &= [L(X);\,L(\hat{X}_b);\,M_b].
\end{align}
Including \(L(X)\) in both streams preserves evidence that may be lost or
misassigned during separation. Each input frame contains \(3F\) features.

We use the masks to guide, rather than hard-restrict, temporal attention.
Let \(A_v=\widetilde{M}_v\) and \(A_b=M_b\). For stream
\(c\in\{v,b\}\), its frame-level confidence is
\begin{equation}
    a_c(t)=
    \operatorname{clip}\!\left(
        \frac{1}{F}\sum_{f=1}^{F}A_c(f,t),
        10^{-4},
        1
    \right).
\end{equation}
The attention bias between non-CLS tokens is
\begin{equation}
    B_c(t,t')
    =
    \alpha\log\sqrt{a_c(t)a_c(t')},
    \qquad \alpha=2.
\end{equation}
Biases involving the CLS token are set to zero, allowing global evidence
aggregation even from low-confidence regions. Each stream is encoded by four
Transformer blocks with hidden dimension \(d=192\) and six attention heads. 

Let \(h_c^{\mathrm{cls}}\) and \(H_c\) denote the resulting CLS embedding
and frame tokens, respectively. A single bidirectional cross-attention step
allows each stream to query complementary evidence from the other:
\begin{align}
    q_{v\leftarrow b}
    &=
    \operatorname{MHA}_v
    \left(h_v^{\mathrm{cls}},H_b,H_b\right),
    \\
    q_{b\leftarrow v}
    &=
    \operatorname{MHA}_b
    \left(h_b^{\mathrm{cls}},H_v,H_v\right).
\end{align}
The final stream representations are
\begin{align}
    r_v^{\mathrm{spec}}
    &=
    \operatorname{LN}
    \left([h_v^{\mathrm{cls}};\,q_{v\leftarrow b}]\right),
    \\
    r_b^{\mathrm{spec}}
    &=
    \operatorname{LN}
    \left([h_b^{\mathrm{cls}};\,q_{b\leftarrow v}]\right).
\end{align}

We additionally retain a direct summary of the pre-cross-attention voice
tokens:
\begin{equation}
    u_v =
    \mathcal{A}_{\mathrm{tok}}
    \left(
        [\operatorname{Mean}_t(H_v);
         \operatorname{Max}_t(H_v)]
    \right),
\end{equation}
where the final layer of \(\mathcal{A}_{\mathrm{tok}}\) is initialized to
zero. This path preserves voice-local evidence while allowing cross-stream
interaction to be introduced gradually. All three outputs, \(r_v^{\mathrm{spec}}, r_b^{\mathrm{spec}},\) and \(u_v\), are 384-dimensional vectors.

\subsection{Complementary Waveform Evidence}
\label{sec:waveform_evidence}

The spectral streams expose component-specific evidence, but the STFT and
source separator may attenuate phase, temporal, and low-level synthesis
artifacts. TruMix therefore analyzes the original mixture through two
complementary waveform encoders.

The contextual branch aims to capture long-range inconsistencies. It employs a pretrained WavLM model~\cite{chen2022wavlm} with a seven-layer convolutional front end and 12 Transformer blocks. We average the outputs of its final four blocks, pool the result over time, and
project the resulting 768-dimensional embedding into the detector's
384-dimensional feature space. To retain the pretrained speech
representation while adapting its high-level features to forensic cues, we
fine-tune only the final two Transformer blocks.

The artifact branch aims to preserve localized waveform traces. It uses a frozen pretrained waveform encoder \cite{baird2025trueear} with parameterized sinc filters, temporal encoding,
and residual aggregation. A trainable adapter maps its output to 384
dimensions.

\subsection{Structured Prediction}
\label{sec:structured_prediction}

A flat eight-class classifier treats recording classes as independent outcomes and learns each class only from examples assigned to that class. This formulation does not reflect our threat model in which an adversary composes a
recording from parts: a synthetic voice may be spliced over authentic
ambient audio to launder its provenance, or an authentic voice placed
over a generated background to defeat environment-based verification.
Therefore, we predict the underlying attributes and assemble the class
distribution from them, so that a synthetic voice supervises the same
head regardless of the background it is paired with.

Let \(r_v^{\mathrm{spec}}\) and \(r_b^{\mathrm{spec}}\) denote the spectral representations of the voice and background components, respectively. Let \(u_v\) denote the voice-token summary, \(r_{\mathrm{ctx}}\) the contextual waveform representation, and \(r_w\) the detector-oriented waveform-artifact representation. We construct the component embeddings as
\begin{align}
    r_v &=
    r_v^{\mathrm{spec}} + u_v + r_w
    + \sigma(g)r_{\mathrm{ctx}},
    \label{eq:voice_fusion}\\
    r_b &= r_b^{\mathrm{spec}},
    \label{eq:background_fusion}
\end{align}
where \(g\) is a learned scalar. We add the artifact representation \(r_w\) directly to the voice embedding because it captures detector-specific evidence, while the broader contextual representation \(r_{\mathrm{ctx}}\) is modulated by the learned gate \(\sigma(g)\). Restricting mixture-level waveform evidence to \(r_v\) prevents it from contaminating the background-specific semantics of \(r_b\).

We represent each recording class using two attributes for each component: presence and authenticity. For a component \(c \in \{v,b\}\), where \(v\) denotes voice and \(b\) denotes background, let $p_c\in\{0,1\}$ be its presence indicator, with \(p_c=1\) if the component is present and \(p_c=0\) otherwise. When \(p_c=1\), let $a_c$$\in\{0,1\}$ denote its authenticity label, with 0 means authentic and 1 means synthetic. 

Each recording class \(y\) is specified by the tuple $\left(p_v^y, a_v^y, p_b^y, a_b^y\right)$, where \(y\) denotes the attribute value associated with class \(y\).  Four binary prediction heads operate on the component embeddings, and each head produces two logits, \(\ell_h \in \mathbb{R}^2\), where \(h \in \{vp,va,bp,ba\}\). We define $q_h(k) = \operatorname{softmax}(\ell_h)_k$, $k \in \{0,1\}$, which is the probability that head \(h\) assigns to outcome \(k\).


For an input recording \(x\), we assign class \(y\) the score \begin{align} s_y &= \log q_{vp}\!\left(p_v^y\right) + \log q_{bp}\!\left(p_b^y\right) \nonumber\\ &\quad + p_v^y \log q_{va}\!\left(a_v^y\right) + p_b^y \log q_{ba}\!\left(a_b^y\right). \label{eq:score} \end{align} 
The presence indicators \(p_v^y\) and \(p_b^y\) mask the corresponding authenticity terms. Consequently, an authenticity head contributes to the class score only when its component is present. The deployed prediction is
\[
P_{\mathrm{final}}(y\mid x)
=
\operatorname{softmax}
([s_1,\ldots,s_8])_y.
\]

Finally, we introduce an auxiliary direct-classification head that predicts the eight-way recording class from the concatenated component embeddings: \begin{equation} \ell^{\mathrm{dir}} = f_{\mathrm{dir}} \left( [r_v; r_b] \right) \in \mathbb{R}^{8}, \label{eq:direct_head} \end{equation} where \([\cdot;\cdot]\) denotes vector concatenation. This head provides auxiliary supervision during training but is discarded at inference time. 

\subsection{Training Objective}
\label{sec:training_objective}

We jointly optimize recording-level classification and component-level
attributes:
\begin{equation}
\begin{aligned}
    \mathcal{L} ={}&
    \mathcal{L}_{8}^{\mathrm{fact}}
    +0.25\mathcal{L}_{8}^{\mathrm{dir}}
    +0.75(\mathcal{L}_{va}+\mathcal{L}_{ba})\\
    &+0.35(\mathcal{L}_{vp}+\mathcal{L}_{bp}).
    \label{eq:total_loss}
\end{aligned}
\end{equation}
The factorized scores and auxiliary direct logits use cross-entropy with
label smoothing \(\epsilon=0.05\):
\begin{equation}
    \mathcal{L}_{8}^{\mathrm{fact}}
    =\operatorname{CE}_{0.05}(s,y),
    \qquad
    \mathcal{L}_{8}^{\mathrm{dir}}
    =\operatorname{CE}_{0.05}(\ell^{\mathrm{dir}},y).
\end{equation}
Label smoothing is not applied to the component-level losses.

Because presence is defined for every recording, each presence head uses
ordinary cross-entropy over the full minibatch:
\begin{equation}
    \mathcal{L}_{cp}
    =
    \frac{1}{N}\sum_{i=1}^{N}
    \operatorname{CE}
    \left(\ell_{cp}^{(i)},p_c^{(i)}\right),
    \qquad c\in\{v,b\}.
    \label{eq:presence_loss}
\end{equation}

Synthetic status is undefined when a component is absent. We therefore use
presence-masked focal cross-entropy with \(\gamma=1.5\). Let
\(q_{ca}^{(i)}\) denote the probability assigned to the ground-truth status
of component \(c\) in sample \(i\):
\begin{equation}
    \mathcal{L}_{ca}
    =
    \frac{
        \displaystyle\sum_{i=1}^{N}
        p_c^{(i)}
        \left[
            -\left(1-q_{ca}^{(i)}\right)^{1.5}
            \log q_{ca}^{(i)}
        \right]
    }{
        \displaystyle\max\left(1,\sum_{i=1}^{N}p_c^{(i)}\right)
    },
    \qquad c\in\{v,b\}.
    \label{eq:status_loss}
\end{equation}
This masking prevents supervision on nonexistent components, while the
factorized recording-level loss jointly calibrates all attribute heads over
the valid class space.

We train TruMix with AdamW using a batch size of 48, weight decay
\(10^{-4}\), and no learning-rate scheduler. The retained configuration uses
a learning rate of \(5\times10^{-6}\) for the trainable detector parameters,
\(5\times10^{-7}\) for the final two WavLM Transformer blocks, and
\(2\times10^{-4}\) for the zero-initialized voice-token mean-maximum readout.  This differential learning-rate schedule limits disruption to pretrained representations while allowing the new component-reasoning layers to adapt rapidly. 


\subsection{MixedTruth Dataset}

We curate MixedTruth, an eight-category mixture dataset spanning five voice
domains: ASVspoof 2021 LA and DF~\cite{yamagishi2021asvspoof}, ASVspoof
5~\cite{wang2024asvspoof}, LJSpeech/WaveFake
~\cite{ljspeech17,frank2021wavefake}, and
MiniLibriSpeech/DECtalk~\cite{panayotov2015librispeech,klatt1982klattalk}.
Real and synthetic environmental backgrounds are drawn from
EnvSDD~\cite{yin2025envsdd}. All MixedTruth partitions use nonoverlapping
source groups. We preserve official source partitions where they are
available. For evaluation-only tracks, such as ASVspoof 2021 LA and DF, we
construct deterministic source-group-disjoint partitions before mixture
synthesis. Therefore, no source clip is shared across the MixedTruth
training and test partitions. 

There are total 6,011,571 clips in MixedTruth, 80\% for training and 20\% for testing. All waveforms are mono at 16~kHz. For examples containing both voice and
background, we use balanced and voice-dominant mixing conditions. All
full-coverage mixtures with both components use nominal target
speech-to-background RMS ratios of approximately 3.19 and 11.74~dB. The same
targets are used for all two-component 2-s held-out mixtures and for the 4-s
held-out mixtures from ASVspoof 2021 LA/DF and ASVspoof 5. The two-component
4-s held-out WaveFake and DECtalk mixtures instead use target ratios of 0 and
20~dB. The 2-s and 4-s datasets are synthesized independently from the
original sources under the same source-group boundaries. 2-s examples are not
crops of 4-s mixtures.

\section{Experiments} \label{sec:exp}

In the experiments, we first evaluate the component-level detection performance of the TruMix. Then, we compare TruMix with 10 state-of-the-art binary synthetic speech detectors. We also provide the ablation analysis and robustness evaluation. For the component-level detection, we treat each class as a one-vs-rest classification task and report the area under the receiver operating characteristic curve (AUC) and equal error rate (EER), computed by varying the decision threshold. To characterize class-specific performance, we also report per-class recall, defined as the fraction of samples from each class that are correctly classified. For ablation analysis, we also report overall accuracy and macro-averaged F1 score. Finally, we quantify robustness using five metrics defined later in the corresponding section.

All experiments were conducted on a workstation equipped with an NVIDIA GeForce RTX 5090 GPU (32 GB of VRAM) and 32 GB of system memory.

\subsection{Component-Level Detection Performance}

\begin{table*}[!t]
\centering
\caption{Component-Level Detection Performance}
\label{tab:eight_category_results}
\footnotesize
\setlength{\tabcolsep}{3.5pt}
\begin{tabular}{lrrrrrr}
\toprule
& \multicolumn{3}{c}{WavLM Base+ with 8-class head}
& \multicolumn{3}{c}{{\bf TruMix}} \\
\cmidrule(lr){2-4}\cmidrule(lr){5-7}
Category & Class accuracy & One-vs-Rest AUC & One-vs-Rest EER $\downarrow$
& Class accuracy & One-vs-Rest AUC & One-vs-Rest EER $\downarrow$ \\
\midrule
RSRB
& 53.627\% & 91.357\% & 16.037\% & \textbf{78.813\%} & \textbf{97.047\%} & \textbf{8.893\%} \\
RSSB
& 42.336\% & 90.878\% & 16.402\% & \textbf{82.960\%} & \textbf{97.811\%} & \textbf{7.337\%} \\
SSRB
& 51.336\% & 90.849\% & 16.517\% & \textbf{78.984\%} & \textbf{97.346\%} & \textbf{8.311\%} \\
SSSB
& 46.189\% & 90.370\% & 16.923\% & \textbf{83.476\%} & \textbf{97.856\%} & \textbf{7.182\%} \\
RS
& 88.636\% & 98.989\% & 4.377\% & \textbf{94.116\%} & \textbf{99.521\%} & \textbf{2.811\%} \\
SS
& 88.293\% & 98.728\% & 4.987\% & \textbf{93.691\%} & \textbf{99.610\%} & \textbf{2.680\%} \\
RB
& 67.927\% & 95.781\% & 10.618\% & \textbf{88.800\%} & \textbf{99.259\%} & \textbf{2.998\%} \\
SB
& 68.780\% & 95.904\% & 9.909\% & \textbf{94.624\%} & \textbf{99.658\%} & \textbf{2.384\%} \\
\midrule
Average (RSRB,RSSB,SSRB, SSSB)
& 48.37\% & 90.86\% &16.47\% & \textbf{81.06\%} & \textbf{97.52\%} & \textbf{7.93\%} \\
Average (all)
&63.39\%  & 94.11\% & 11.97\%& \textbf{86.93\%} & \textbf{98.51\%} & \textbf{5.33\%} \\
\bottomrule
\end{tabular}
\end{table*}

In the first round of experiments, we evaluate the component-level detection performance of TruMix. 
	The evaluation covers eight combinations of speech and background components: real speech in a real background (RSRB), real speech with a synthetic background (RSSB), synthetic speech with a real background (SSRB), synthetic speech with a synthetic background (SSSB), real or synthetic speech alone (RS/SS), and real or synthetic background alone (RB/SB).

For comparison, we fine-tune WavLM Base+ \cite{chen2022wavlm} with an eight-class classification head using the same training dataset as TruMix.  	We select this primary baseline because it provides a strong yet controlled test of TruMix's architectural contribution. WavLM is a widely adopted self-supervised speech encoder trained to capture both linguistic content and non-semantic acoustic characteristics, including speaker, channel, and environmental information. Its pretraining objective additionally exposes the model to corrupted and mixed utterances, which makes it particularly relevant to recordings containing both foreground speech and background audio. The Base configuration contains a convolutional feature encoder followed by 12 Transformer blocks with a 768-dimensional hidden representation, providing substantial capacity for learning synthetic-audio cues directly from raw waveforms. The WavLM has reported strong performance across a broad range of speech-processing tasks.

Table \ref{tab:eight_category_results} presents the performance of TruMix compared to the baseline. Observe that TruMix consistently outperforms the baseline across all eight categories and all three metrics. Averaged across all  categories, TruMix improves class accuracy from 63.39\% to 86.93\%, a gain of 23.54 percentage points. It also increases the macro-average AUC from 94.11\% to 98.51\% and reduces the macro-average EER from 11.97\% to 5.32\%, corresponding to a 55.5\% relative reduction. Importantly, these gains are not driven by only one component type.  TruMix improves every category in accuracy and reduces EER in every case.

The advantage of TruMix is most pronounced for mixed-content recordings, which are the most challenging and security-relevant cases. Specifically, across RSRB, RSSB, SSRB, and SSSB, the baseline obtains an average class accuracy of only 48.37\%, despite achieving a relatively high average AUC of 90.86\%. This discrepancy indicates that the baseline learns some discriminative evidence but struggles to assign the correct joint label when speech and background have different provenance. In contrast, TruMix raises the average accuracy on these four classes to 81.06\%, increases AUC to 97.52\%, and reduces EER from 16.47\% to 7.93\%. The improvement indicates that explicitly modeling the two audio components is particularly important when authentic and synthetic evidence coexist within the same recording.  

The largest class-accuracy gain occurs for RSSB, for which TruMix improves accuracy from 42.34\% to 82.96\%. TruMix similarly improves SSSB by 37.29 points and SSRB by 27.65 points. These results expose a fundamental limitation of the single-path WavLM classifier when a shared representation must simultaneously encode speech presence, background presence, and the authenticity of each component. Dominant cues from one component can therefore mask weaker or contradictory evidence from the other. For example, natural speech may dominate the representation of RSSB and cause the synthetic background to be overlooked. TruMix mitigates this interference by routing estimated speech and background through separate spectral branches and allowing the branches to exchange evidence through bidirectional cross-attention.
	
	
	The observed gains align with TruMix's multi-path design. First, the speech-separation front end produces explicit voice and background streams, which enables the dual spectral Transformer path to examine component-specific spectral evidence rather than forcing a single encoder to disentangle it implicitly. Mask-guided attention emphasizes time–frequency regions assigned to each component, while speech-to-background and background-to-speech cross-attention capture inconsistencies between the two streams. Such inconsistencies are particularly informative for hybrid cases such as RSSB and SSRB. Second, the raw-waveform contextual path preserves global semantic, temporal, and acoustic information that may be weakened by source separation or spectral projection. Third, the waveform-artifact path is designed to capture localized generation and signal-processing traces that may not be represented reliably by contextual embeddings. The four resulting representations, i.e., voice spectral, background spectral, contextual waveform, and waveform artifact, provide complementary views of the same recording. 
	
	
In summary, the results demonstrate that detecting partially synthetic audio is not adequately addressed by attaching a multiclass head to a general-purpose speech encoder. TruMix's multi-path architecture design provides a more effective way in component-level reasoning.

\subsection{Comparisons with State-of-the-art \\Synthetic Voice Detectors}

We compare TruMix with 10 state-of-the-art synthetic voice detectors: AASIST \cite{jung2022aasist}, AASIST-L~\cite{jung2022aasist}, RawNet2~\cite{tak2021end}, AASIST2
~\cite{zhang2024improving}, AASIST3~\cite{borodin2024aasist3}, Trident of Poseidon
~\cite{doan2024trident}, XLS-R+SLS~\cite{zhang2024audio}, Nes2Net-X
~\cite{liu2025nes2net}, ALDEN~\cite{xu2025alden}, and TrueEar~\cite{baird2025trueear}. These detectors are reviewed in Section \ref{sec:related} while  Table \ref{tab:detector_architecture_complexity} summarizes their architectures and model sizes.

 Because the baseline detectors produce only a binary real-versus-synthetic decision, we map the eight MixedTruth classes to binary labels. An example
is labeled \emph{synthetic} if either its voice or background component is synthetic. Thus, SS, SB, SSRB, SSSB, and RSSB are mapped to the synthetic class, whereas RS, RSRB, and RB are mapped to the real class. For a direct
comparison, we apply the same mapping to TruMix's eight-class output. In
particular, we compute the binary synthetic score by summing the predicted
probabilities of the five synthetic-containing classes:
\begin{equation}
	s_{\mathrm{syn}} =
	p_{\mathrm{SS}} + p_{\mathrm{SB}} + p_{\mathrm{SSRB}}
	+ p_{\mathrm{SSSB}} + p_{\mathrm{RSSB}}.
\end{equation}
This probability aggregation avoids introducing a detector-specific decision threshold when computing AUC and EER.

\subsubsection{Comparison with Synthetic-Voice Detectors in Their Original Form}

In this round of experiments, each baseline detector receives the original unseparated waveform from the MixedTruth test set. We evaluate the publicly released models using their published configurations and decision scores. Results are reported separately for clean speech without a background component
(RS/SS) and for all background-containing examples
(RSRB, RSSB, SSRB, SSSB, RB, and SB). The clean-speech partition includes voice domains represented in the baseline detectors' training data. In contrast, the background-containing partition is substantially more challenging as background audio can obscure artifacts introduced by speech generation while synthetic background content may provide independent evidence of manipulation.

\begin{table}[!ht]
	\centering
	\caption{Comparison with state-of-the-art synthetic-voice detectors. ``+BG'' denotes all examples containing a background component. Best results are shown in bold.}
	\label{tab:binary_detector_comparison}
	\footnotesize
	\setlength{\tabcolsep}{5pt}
	\begin{tabular}{lrrrr}
		\toprule
		Detector & RS/SS AUC & RS/SS EER $\downarrow$ & +BG AUC & +BG EER $\downarrow$ \\
		\midrule
		TrueEar & 73.094\% & 32.316\% & 57.847\% & 44.513\% \\
		AASIST & 79.703\% & 28.533\% & 64.394\% & 39.267\% \\
		AASIST-L & 79.889\% & 28.229\% & 59.502\% & 43.762\% \\
		RawNet2 & 76.017\% & 31.373\% & 60.578\% & 42.236\% \\
		AASIST2 & 90.361\% & 20.773\% & 74.341\% & 32.142\% \\
		AASIST3 & 76.409\% & 31.444\% & 61.680\% & 41.704\% \\
		Trident  & 93.681\% & 12.344\% & 77.588\% & 27.798\% \\
		XLS-R & 94.322\% & 11.247\% & 75.923\% & 30.193\% \\
		Nes2Net-X & 94.485\% & 12.338\% & 78.086\% & 27.864\% \\
		ALDEN & 90.585\% & 16.944\% & 72.750\% & 32.036\% \\
		\midrule
		{\bf TruMix} & \textbf{99.000\%} & \textbf{4.860\%}
		& \textbf{95.622\%} & \textbf{9.963\%} \\
		\bottomrule
	\end{tabular}
\end{table}

As shown in Table~\ref{tab:binary_detector_comparison}, TruMix consistently outperforms all ten binary detectors under both evaluation conditions. On the RS/SS subset, TruMix achieves an AUC of 99.000\% and an EER of 4.860\%. The advantage of TruMix becomes substantially larger on background-containing audio. TruMix obtains 95.622\% AUC and 9.963\% EER, whereas the best baseline AUC is 78.086\%, achieved by Nes2Net-X, and the lowest baseline EER is 27.798\%, achieved by Trident of Poseidon. TruMix therefore improves AUC by 17.536 percentage points and reduces EER by
17.835 percentage points, corresponding to a 64.2\% relative reduction in EER.
We can also observe that all conventional detectors experience a significant degradation when background components are introduced. These results indicate that detectors trained primarily to identify speech-generation artifacts do not reliably distinguish those artifacts
when they are attenuated or confounded by background audio.


TruMix’s improved robustness stems from its mixture-aware design. Rather than representing the entire waveform with a single embedding, TruMix integrates separated voice and background spectral representations with contextual features from the original waveform and waveform-level artifact features. This design substantially improves detection in mixed-content audio while also enhancing performance in conventional clean-speech settings, as demonstrated by the RS/SS results.



\subsubsection{Comparison with Separation-Augmented Synthetic-Voice Detectors} \label{sec:separator}

A key question is whether the challenge of detecting synthetic content in
mixed audio can be addressed simply by placing a speech-separation front end
before an existing detector. To evaluate this hypothesis, we construct a
separation-augmented version of each baseline by using the state-of-the-art
Conv-TasNet~\cite{luo2019conv} to decompose each mixed waveform into an
estimated voice signal and an estimated background signal. The same baseline
detector is then applied independently to both separated components. Neither
the separator nor the baseline detectors are retrained or fine-tuned on
MixedTruth, allowing us to isolate the effect of separation at inference time.

We evaluate the original and separation-augmented detectors on the same
balanced subset of MixedTruth, comprising 1,000 RSRB examples and 1,000 SSSB
examples from each domain. This setting requires distinguishing fully authentic mixtures from
mixtures in which both the voice and background are synthetic. For the
separation-augmented detectors, a sample is classified as real only when both
separated components are classified as real; it is classified as synthetic
if either component is classified as synthetic. To compute threshold-independent
metrics, we fuse the two component-level synthetic scores using
\begin{equation}
    s_{\mathrm{sep}} = \max(s_{\mathrm{speech}},s_{\mathrm{background}}),
\end{equation}
which is the score-level counterpart of the above logical decision rule.
We report AUC and EER for each detector when applied directly to the original
mixture (``Original'') and when used in the separation-augmented cascade
(``Sep.'').

\begin{table}[!t]
    \centering
    \caption{Effect of augmenting existing synthetic-voice detectors with a
    voice-separation front end.}
    \label{tab:separation_cascade}
    \footnotesize
    \setlength{\tabcolsep}{4pt}
    \begin{tabular}{lrrrr}
        \toprule
        Detector &
        Original AUC &
        Original EER $\downarrow$ &
        Sep. AUC &
        Sep. EER $\downarrow$ \\
        \midrule
        TrueEar             & 57.97\% & 45.59\% & 65.57\% & 39.03\% \\
        AASIST              & 70.22\% & 33.19\% & 69.72\% & 32.97\% \\
        AASIST-L            & 67.37\% & 36.44\% & 66.05\% & 37.41\% \\
        RawNet2             & 64.64\% & 38.94\% & 68.73\% & 34.34\% \\
        AASIST2             & 87.71\% & 20.62\% & 79.78\% & 28.93\% \\
        AASIST3             & 67.42\% & 37.53\% & 59.23\% & 43.07\% \\
        Trident             & 89.78\% & 17.93\% & 78.12\% & 28.08\% \\
        XLS-R+SLS           & 88.84\% & 18.18\%
                            & \underline{85.77\%} & \underline{23.56\%} \\
        Nes2Net-X           & \underline{90.18\%} & \underline{16.79\%}
                            & 76.65\% & 31.64\% \\
        ALDEN               & 82.83\% & 24.71\% & 75.39\% & 30.97\% \\
        \midrule
        \textbf{TruMix}     & \textbf{98.23\%} & \textbf{6.54\%}
                            & \textbf{94.80\%} & \textbf{11.83\%} \\
        \bottomrule
    \end{tabular}
\end{table}

Table~\ref{tab:separation_cascade} shows that separation alone does not
provide a reliable solution to mixed-audio detection. Although the separation
front end improves several weaker detectors, it degrades seven of the ten
detectors in EER and eight of the ten detectors in AUC. More importantly,
separation degrades all five strongest original-mixture baselines, i.e., AASIST2,
Trident of Poseidon, XLS-R+SLS, Nes2Net-X, and ALDEN, on both metrics.
For example, Nes2Net-X's AUC decreases from 90.18\% to 76.65\%, while its
EER increases from 16.79\% to 31.64\%. 


On the same evaluation subset, TruMix achieves 98.23\% AUC and 6.54\% EER
on the original mixtures. Under the same separation-augmented protocol, it
achieves 94.80\% AUC and 11.83\% EER, exceeding the best separation-augmented
baseline. 


These results reveal an important limitation of a naive separate-then-detect cascade. Existing synthetic-voice detectors are typically trained on natural or codec-processed speech waveforms rather than separator outputs. Separation can suppress or distort the subtle generation
artifacts on which these detectors rely, while introducing separator-specific artifacts that shift the input distribution.

In contrast, separation is an integrated component of TruMix rather than an
independent preprocessing step. TruMix jointly learns from the separated
voice and background spectra, the original mixed waveform, and waveform-level
artifact features. Its mask-guided attention and bidirectional cross-attention
also preserve and model interactions between the two components. The results
therefore support our central design choice. Robust mixed-audio detection
requires mixture-aware feature learning and joint evidence fusion, rather
than merely appending a separator to an existing binary detector.

\subsection{Ablation Analysis}

\begin{table*}[!t]
\centering
\caption{Ablation analysis for TruMix}
\label{tab:cumulative_ablation_paper}
\footnotesize
\setlength{\tabcolsep}{3.5pt}
\begin{tabular}{lllll}
\toprule
Configuration & Accuracy & Macro F1 & Voice EER $\downarrow$ & BG EER $\downarrow$ \\
\midrule
(1) WavLM Base+ with 8-class multi-head
& 63.390\% & 63.096\% & 12.904\% & 38.581\% \\
(2) + separated voice/background dual paths
& 85.827\% (+22.436) & 85.812\% (+22.716)
& 9.325\% (-3.579) & 7.407\% (-31.173) \\
(3) + waveform-artifact Path D
& 85.872\% (+22.482) & 85.855\% (+22.759)
& 9.206\% (-3.699) & 7.420\% (-31.161) \\
(4) + structured multi-head fusion
& 85.783\% (+22.392) & 85.761\% (+22.665)
& 8.953\% (-3.951) & 7.441\% (-31.139) \\
(5) Complete model
& \textbf{86.933\% (+23.543)} & \textbf{86.917\% (+23.821)}
& \textbf{7.419\% (-5.485)} & \textbf{7.174\% (-31.407)} \\
\bottomrule
\end{tabular}
\end{table*}
	
	We add TruMix's principal components cumulatively to a pretrained 12-block WavLM Base+ encoder with an eight-class multi-head classifier. All variants share the same 360{,}000-example combined 2-s and 4-s test set and follow an identical training and evaluation protocol. 	Table~\ref{tab:cumulative_ablation_paper} reports the performance of each variant.

In configuration (1), the fine-tuned WavLM Base+ encoder is the backbone of current state-of-the-art synthetic-speech detectors, and it reflects the detection capability an analyst would have in practice. Its voice EER of 12.904\% confirms that this backbone remains competent at the task it was designed for. Its background EER of 38.581\%, against a 50\% chance floor, shows that this competence does not transfer. A representation trained to detect vocoder artifacts in speech carries almost no signal about whether the surrounding acoustic environment was synthesized. The failure is structural rather than a matter of capacity, and it is exactly the failure an adversary exploits by leaving the voice authentic and fabricating the room.

In configuration (2), the dual spectral Transformer paths (Path B in
Figure~\ref{fig:trumix}) provide the largest single performance gain: accuracy
and macro F1 increase by 22.44 and 22.72 percentage points, respectively,
while background EER drops sharply from 38.581\% to 7.407\%. Voice EER also
decreases from 12.904\% to 9.325\%. These results demonstrate that a shared
representation of the original mixture is insufficient for component-level
forensic analysis. By explicitly modeling the speech and background streams,
TruMix can expose component-specific manipulation cues that might otherwise
be obscured by the other source. Moreover, these gains cannot be attributed
to source separation alone. As shown in Section~\ref{sec:separator}, applying
the same frozen separator to existing detectors and classifying the separated
streams independently degrades the overall performance of seven of the ten
detectors and yields only modest improvements for the remaining three. The
advantage of TruMix is therefore structural. A conventional
separate-then-detect pipeline treats the separator's estimates as definitive
inputs, making downstream decisions vulnerable to separation errors and
artifacts. In contrast, TruMix treats the separated streams as complementary
conditional evidence and jointly reasons over them rather than assuming that
they are ground-truth sources.

In configuration (3), adding the waveform-artifact path yields a smaller but consistent improvement in overall classification and voice authentication, increasing accuracy to 85.872\% and reducing voice EER to 9.206\%. Its limited effect on background EER suggests that this path primarily captures low-level synthesis traces associated with generated speech, whereas background authenticity is already well represented by the separated spectral stream. 
    
In configuration (4), structured multi-head fusion further reduces voice EER to 8.953\%, despite a slight decrease in eight-class accuracy and macro F1. This tradeoff indicates that factorizing the prediction into presence and authenticity decisions improves component-specific discrimination, even when the resulting decisions do not always form the correct joint class.
	
	The complete model achieves the best result for every metric. Relative to structured fusion alone, it raises accuracy and macro F1 by 1.150 and 1.156 percentage points and reduces voice and background EER by 1.534 and 0.267 points, respectively. Relative to the baseline, TruMix improves accuracy by 23.543 points and macro F1 by 23.821 points, while reducing voice and background EER by 5.485 and 31.407 points. The particularly strong final improvement in voice EER shows that contextual evidence from the original mixed waveform complements the source-specific and low-level forensic paths. Separation exposes localized manipulation traces, while the intact mixture preserves cross-component relationships and acoustic consistency. Overall, the ablation demonstrates that TruMix's performance does not arise from a stronger classifier alone. Its gains result from combining source-aware spectral analysis, waveform-level artifact detection, mixture context, and structured joint reasoning, with each view addressing a distinct weakness of component-level audio authentication.

\subsection{Robustness Evaluation}
\label{sec:robustness}

We evaluate the robustness of {TruMix} under the four attack scenarios as described in Section \ref{sec:threat}.  We consider an evasion adversary whose objective is to conceal
synthetic content rather than merely cause an arbitrary classification error.
In particular, the adversary attempts to cause an input containing at least one
synthetic component to receive the corresponding fully authentic verdict:
\[
\tau(y)=
\begin{cases}
\mathrm{RSRB}, & y\in\{\mathrm{RSSB},\mathrm{SSRB},\mathrm{SSSB}\},\\
\mathrm{RS},   & y=\mathrm{SS}\\
\end{cases}
\]
For example, classifying an SSRB recording as RSSB is an attribution error but
does not satisfy the adversary's primary objective because the recording is
still identified as containing synthetic content and will trigger a system alert.


    

It is worth noting that a change in the detector's output is not by itself a successful real-world
attack. The resulting recording must continue to convey the intended speech and
speaker identity since an unintelligible or severely corrupted recording would be unlikely to deceive a human or an AI agent. We therefore adopt the following  metrics to provide a comprehensive measurement. 

\vspace{5pt}
\noindent$\bullet$ \textbf{Targeted Misclassification Rate (TMR).}
   Let     $\mathcal{O}$ denote the set of original speech related examples that contain the synthetic components and  are classified correctly by TruMix before attack. 
    \[
       \mathcal{O} =
       \left\{x_i :
       y_i\in\{\mathrm{RSSB},\mathrm{SSRB},\mathrm{SSSB},
                    \mathrm{SS}\}
       \land f(x_i)=y_i
       \right\}.
    \]
        Let $f(\cdot)$ denote the predicted {TruMix} label. Let $x_i$ and
    $x_i'$ denote the original recording in $\mathcal{O}$ and its perturbed version, respectively.  The targeted misclassification rate  (TMR) is calculated as follows:
    \[
       \mathrm{TMR}
       =
       \frac{1}{|\mathcal{O}|}
       \sum_{x_i\in\mathcal{O}}
       \mathbbm{1}\!\left[f(x_i')=\tau(y_i)\right].
    \]
The TMR measures the percentage of perturbed inputs being misclassified as the target label. This metric considers only initially correctly classified inputs, which ensures the isolation of  the original detection error from misclassifications induced by the attack. 
    
\vspace{5pt}
\noindent$\bullet$  \textbf{Content Preservation Rate (CPR).}
A higher TMR does not directly imply a successful attack if the content of speech has been distorted from the original one. To measure the content preservation rate (CPR), we  transcribe the attacked audio using  an automatic speech-recognition system (Wav2Vec 2.0\cite{baevski2020wav2vec}),  compare the result with the     ground-truth transcript, and then  compute the word error rate. Specifically, let $S_i$, $D_i$, and $I_i$ be the numbers of word
    substitutions, deletions, and insertions, respectively,  and let $N_i$ be the number of words  in the reference transcript, then the content preservation rate (CPR) is defined as follows: 
    \[
       \mathrm{CPR}_i =  \max\left(0,\,
       1-\frac{S_i+D_i+I_i}{N_i}\right).
    \]

\vspace{5pt}
\noindent$\bullet$ \textbf{Speaker Preservation Rate (SPR).}
We apply the frozen TruMix separator to the original and attacked recordings
and extract their speech channels. A pretrained WavLM Base+
speaker-verification model produces $\ell_2$-normalized speaker embeddings. The SPR is calculated as the mean cosine similarity over all launched
original-attacked pairs.


\vspace{5pt}
\noindent$\bullet$  \textbf{Signal-to-noise ratio (SNR).}
We use the commonly adopted signal-to-noise ratio (SNR) to assess the quality of perturbed audio. As a practical reference, speech with an SNR above 20 dB is generally considered clear and readily intelligible, while an SNR above 30 dB indicates relatively minor distortion. Because SNR alone may not fully capture speech intelligibility, we report it along with other metrics described in this section.

\vspace{5pt}
\noindent$\bullet$ {\bf Valid Evasion Rate (VER).} Finally, VER integrates the first four objective metrics to quantify the proportion of attacks that both achieve the intended target label and preserve the deceptive utility of the original recording. Specifically, an attack is considered a valid evasion if the perturbed recording (1) is classified as the target label, (2) achieves content- and speaker-preservation rates above 80\%, and (3) has an SNR of at least 20 dB. The valid evasion rate is calculated as the number of attacks satisfying all these conditions divided by the total number of attacks launched.

\subsubsection{Model-Agnostic Attacks}
We evaluate whether TruMix remains reliable under model-agnostic transformations as defined in Section \ref{sec:threat}. We apply three common classes of transformations to  a fixed set of 1,000 audio recordings: (i) changes in background intensity, (ii) reverberation, and (iii) lossy audio compression. We also evaluate randomly composed transformations to approximate less predictable post-processing pipelines. All conditions use the same underlying recordings, enabling direct comparison across transformations. We report the percentage of malicious recordings misclassified by TruMix and  the strongest baseline Nes2Net-X. Lower values indicate greater robustness.

In Table~\ref{tab:model_agnostic_teacher_attack}, the ``Original'' row reports performance on unmodified recordings and serves as the reference point for the subsequent transformations. Consistent with the results in the previous section, TruMix substantially outperforms Nes2Net-X on these unmodified recordings, reducing the misclassification rate from 30.30\% to 4.70\%. As discussed below, this advantage persists across nearly all evaluated transformations.

\begin{table}[!ht]
\centering
\caption{Robustness to Model-Agnostic Audio Transformations. Lower values indicate better robustness.}
\label{tab:model_agnostic_teacher_attack}
\footnotesize
\setlength{\tabcolsep}{3pt}
\begin{tabular}{lrr}
\toprule
\textbf{Transformation}
& \textbf{TMR (TruMix) $\downarrow$}
& \textbf{TMR (Nes2Net-X) $\downarrow$} \\
\midrule\midrule
Original                              & 4.70\%  & 30.30\% \\
\midrule \midrule
0\% background                        & 0.20\%  & 27.70\% \\
10\% background                       & 4.80\%  & 28.90\% \\
20\% background                       & 7.30\%  & 31.20\% \\
60\% background                       & 8.30\%  & 33.50\% \\
80\% background                       & 1.60\%  & 59.20\% \\
\midrule

Small-room reverb (speech)             & 7.90\%  & 24.30\% \\
Small-room reverb (+BG)           & 4.50\%  & 24.60\% \\
Large-room reverb (speech)             & 22.40\% & 16.00\% \\
Large-room reverb (+BG)           & 16.00\% & 17.50\% \\
Outdoor reverb (speech)                & 7.10\%  & 29.70\% \\
Outdoor reverb (+BG)              & 3.90\%  & 29.40\% \\
\midrule
MP3                                    & 3.20\%  & 28.50\% \\
Opus                                   & 4.30\%  & 26.10\% \\
AAC                                    & 3.20\%  & 28.80\% \\
\midrule
Random combination                     & 6.10\%  & 34.50\% \\
\midrule
{\bf Overall Average} & {\bf 6.72\%} & {\bf 29.33\%}  \\
\bottomrule
\end{tabular}
\end{table}

TruMix remains effective under substantial variation in background intensity, with targeted misclassification rates (TMR) ranging from 0.20\% to 8.30\%, compared with 27.70\% to 59.20\% for Nes2Net-X. The difference is most significant at 80\% background intensity  whereby TruMix achieves an error rate of only 1.60\%, whereas the error rate of Nes2Net-X rises sharply to 59.20\%. This contrast is particularly important for mixed-audio forensics because detectors that rely primarily on aggregate speech characteristics may lose reliable voice evidence as the background becomes dominant. By explicitly modeling both components, TruMix can treat prominent background content as an additional source of authenticity evidence rather than merely as interference that obscures speech.

Small-room and outdoor reverberation have relatively limited effects on TruMix. When reverberation is applied to both speech and background, TruMix achieves error rates of 4.50\% and 3.90\% under the small-room and outdoor conditions, respectively, compared with 24.60\% and 29.40\% for Nes2Net-X. Moreover, applying reverberation to both components results in lower error than reverberating speech alone. This finding suggests that joint component analysis provides complementary evidence when reverberation distorts speech-specific cues. Large-room reverberation is the most challenging transformation for TruMix and is the only setting in which Nes2Net-X achieves a lower error rate. One possible reason is that strong long-duration reverberation smears temporal and spectral synthesis artifacts while altering the apparent relationship between speech and its acoustic surroundings, thereby weakening several forensic views simultaneously. Nevertheless, applying large-room reverberation to both components reduces TruMix's error from 22.40\% to 16.00\%, which again highlights the value of modeling cross-component acoustic consistency. It also identifies severe reverberation as an important target for future robustness enhancements.

TruMix is also highly robust to common lossy audio codecs, achieving misclassification rates of 3.20\%, 4.30\%, and 3.20\% under MP3, Opus, and AAC compression, respectively. In contrast, Nes2Net-X remains above 26\% under all three codecs. These results suggest that TruMix does not rely solely on fragile synthesis artifacts that may be suppressed by compression.

Under randomly combined transformations, TruMix misclassifies only 6.10\% of the recordings, compared with 34.50\% for Nes2Net-X. Thus, its complementary forensic paths remain effective even when multiple distortions interact. 

Finally, the last row reports the average misclassification rate across all 15 transformation conditions. TruMix achieves a misclassification rate of 6.72\%, versus 29.33\% for Nes2Net-X. Thus, TruMix's advantage is not limited to a particular post-processing operation or to the original evaluation distribution. Its component-level reasoning improves not only attribution accuracy but also resilience to realistic detector-independent post-processing.   

For these transformation-based attacks, we omit additional metrics, such as content and speaker preservation, because TruMix’s misclassification rate is already very low and the valid evasion rate would only be even lower.

\subsubsection{Decision-based Black-box Attacks}
\label{sec:blackbox_attack}


We use HopSkipJump~\cite{chen2020hopskipjumpattack} to evaluate a
decision-based black-box adversary that observes only TruMix's predicted
class label. The adversary has no access to model parameters, gradients, or
confidence scores. We use the targeted $\ell_\infty$ variant of HopSkipJump,
which searches for an input classified as the exact authentic counterpart
of the source category while progressively reducing its distance from the
original waveform.

For each of the four source categories (RSSB, SSRB, SSSB, and SS), we select
500 recordings that TruMix classifies correctly before attack, yielding
2,000 source recordings. We evaluate the attack across query budgets ranging from 100 to 5,000. The three query-budget conditions are therefore snapshots of the same 2,000 attack trajectories rather than independent attacks, producing 6,000 saved adversarial waveforms. Queries used for target-class initialization are included in the attack budget. Each saved waveform is reloaded and evaluated again before its target prediction and waveform-quality metrics are recorded.

Table~\ref{tab:blackbox} exposes a substantial gap between nominal attack
success and valid evasion. HopSkipJump reaches the target class for nearly
every input with only 100 queries, but just 0.13\% of these waveforms
satisfy all validity constraints. Larger budgets  improve
CPR, SPR, SNR, and VER, whereas TMR remains near 100\%, indicating that
additional queries primarily reduce distortion after target-class
reachability has been established. RSSB is the easiest category to compromise, which is likely because  its real-speech component already matches the
target and the attack primarily needs to alter evidence associated with
background authenticity. SS is the most resistant since an SS-to-RS attack
must reverse the speech-authenticity decision while preserving both
content and speaker identity. Even at 5,000
queries, only 9.75\% of attacks constitute valid
evasions overall and only 0.50\% do so for SS.  These results demonstrate that target misclassification
alone substantially overestimates the practical effectiveness of
decision-based black-box attacks.


\begin{table}[!t]
\centering
\caption{Decision-based black-box attacks against TruMix.}
\label{tab:blackbox}
\footnotesize
\setlength{\tabcolsep}{2.7pt}
\begin{tabular}{@{}llrrrrr@{}}
\toprule
\textbf{Category} &
\textbf{Queries} &
\textbf{TMR} &
\textbf{CPR} &
\textbf{SPR} &
\textbf{SNR (dB)} &
\textbf{VER} \\
\midrule
RSSB & 100 & 100.00\% & 7.28\% & 66.06\% & -0.87 & 0.50\% \\
SSRB & 100 & 100.00\% & 0.85\% & 62.54\% & -1.76 & 0.00\% \\
SSSB & 100 & 100.00\% & 0.90\% & 65.19\% & -1.65 & 0.00\% \\
SS   & 100 & 100.00\% & 0.69\% & 61.76\% & -0.40 & 0.00\% \\
\textbf{Overall} &
\textbf{100} &
\textbf{100.00\%} &
\textbf{2.43\%} &
\textbf{63.89\%} &
\textbf{-1.17} &
\textbf{0.13\%} \\
\midrule
RSSB & 1{,}000 & 100.00\% & 35.77\% & 84.56\% & 10.82 & 18.00\% \\
SSRB & 1{,}000 & 100.00\% & 8.38\%  & 70.65\% & 2.42  & 2.50\% \\
SSSB & 1{,}000 & 100.00\% & 7.03\%  & 72.63\% & 1.89  & 0.50\% \\
SS   & 1{,}000 & 100.00\% & 1.22\%  & 61.78\% & 1.02  & 0.00\% \\
\textbf{Overall} &
\textbf{1{,}000} &
\textbf{100.00\%} &
\textbf{13.10\%} &
\textbf{72.41\%} &
\textbf{4.04} &
\textbf{5.25\%} \\
\midrule
RSSB & 5{,}000 & 99.80\%  & 49.73\% & 92.69\% & 17.74 & 23.50\% \\
SSRB & 5{,}000 & 99.60\%  & 25.51\% & 81.48\% & 8.55  & 10.50\% \\
SSSB & 5{,}000 & 100.00\% & 18.27\% & 81.64\% & 6.82  & 4.50\% \\
SS   & 5{,}000 & 99.80\%  & 4.03\%  & 65.22\% & 2.69  & 0.50\% \\
\textbf{Overall} &
\textbf{5{,}000} &
\textbf{99.80\%} &
\textbf{24.38\%} &
\textbf{80.26\%} &
\textbf{8.95} &
\textbf{9.75\%} \\
\bottomrule
\end{tabular}
\end{table}

\subsubsection{Gray-Box Transfer Attacks}
\label{sec:graybox_attack}


We select the previously introduced WavLM Base+ with an eight-class classification head as a strong surrogate model.  Recall that it is trained using the same training dataset as TruMix, which provides the attacker with a favorable setting in which even the exact training samples are known.  The attacker, however, has no access to TruMix's parameters, or gradients. 


We use a 20-step projected-gradient-descent attack (PGD-20) with an $\ell_\infty$ budget of $\epsilon=0.002$. We select this budget because it yields a nearly $100\%$ TMR in the white-box setting evaluated in the following subsection, thereby providing a strong basis for assessing cross-model transferability.  For each of the four input categories (RSSB, SSRB, SSSB, and SS), we launch 1,000 attacks. The target class is RSRB for RSSB, SSRB, and SSSB inputs, and RS for SS inputs.

Table~\ref{tab:graybox_ver_conditional_cpr_n1000_allpair_spksim_v3} reports both attack effectiveness and perceptual validity. The targeted misclassification rate (TMR) measures how often TruMix predicts the attacker-selected class, without imposing perceptual-quality constraints. To distinguish a label change from a usable voice-evasion example, we also report the valid evasion rate (VER). VER requires simultaneous exact target success,  content-preservation ratio (CPR) above $0.80$, speaker similarity above $0.80$, and signal-to-noise ratio (SNR) of at least $20$\,dB. 

\begin{table}[!ht]
\centering
\caption{Gray-box transfer attacks against TruMix.}
\label{tab:graybox_ver_conditional_cpr_n1000_allpair_spksim_v3}
\footnotesize
\setlength{\tabcolsep}{2.7pt}
\begin{tabular}{@{}lrrrrr@{}}
\toprule
\textbf{Category} &
\textbf{TMR} &
\textbf{CPR} &
\textbf{SPR} &
\textbf{SNR (dB)} &
\textbf{VER} \\
\midrule
RSSB & 0.90\% & 72.00\% & 99.65\% & 33.96 & 0.75\% \\
SSRB & 25.40\% & 46.65\% & 99.14\% & 35.42 & 1.75\% \\
 SSSB & 0.90\% & 0.00\% & 99.18\% & 35.06 & 0.00\% \\
 SS & 23.10\% & 56.66\% & 96.97\% & 35.67 & 2.00\% \\
 \textbf{Overall} & \textbf{12.58\%} & \textbf{51.72\%} &
{98.74\%} & 35.50 & \textbf{1.13\%} \\
\bottomrule
\end{tabular}
\end{table}

As shown in the table, the transferred perturbations achieve an overall TMR of $12.58\%$. Inputs with either no background (SS) or a real background (SSRB) exhibit higher TMRs, at $23.10\%$ and $25.40\%$, respectively. By contrast, transfer is substantially weaker for RSSB and SSSB, with only $0.90\%$ of attacks reaching the specified target. This is likely because  the surrogate model is predominantly speech-oriented which is also the case of existing synthetic speech detectors. As a result, perturbations optimized on the surrogate are not able to transfer to  background-related authenticity. 

After taking into account the  perceptual constraints, the overall VER drops to  only $1.13\%$. In particular, SSRB's $25.40\%$ TMR decreases to a $1.75\%$ VER, while SS decreases from $23.10\%$ to $2.00\%$. These gaps show that target matching alone substantially overestimates the practical effectiveness of the transferred attacks. The average content preservation rate (CPR) is only around 51.72\%. Note that these averages or perceptual metrics should not be interpreted as success-conditioned perceptual quality, since they include all launched attacks and do not guarantee that the same example simultaneously satisfies the target, achieves above 80\% content similarity, above 80\% speaker-similarity, and above 20dB SNR conditions. 

Overall, these results demonstrate limited transferability from the WavLM surrogate under the evaluated gray-box attack. TruMix's heterogeneous evidence paths that separated voice and background spectra, raw-waveform context, and waveform-artifact features, may reduce reliance on the surrogate-specific representation exploited by PGD.

\subsubsection{White-box Attacks}

	Finally, we evaluate robustness under the strongest  white-box threat model in which the attacker has complete knowledge of the detector. We compare TruMix with Nes2Net-X, which we select because it achieves the highest detection rate among the existing detectors evaluated in our study. Because Nes2Net-X is a binary detector, an attack is considered a targeted misclassification when it causes an input from any synthetic category to be classified as real.
	
We evaluate PGD-20 at various budgets. We also employ AutoAttack \cite{croce2020reliable} which is  an ensemble of complementary parameter-free attacks designed to reduce the risk of overestimating robustness due to attack-specific hyperparameter choices or gradient-masking effects. For each detector, attack, and perturbation budget, we launch 1,000 attacks from each of the four source categories, totaling 4,000 attacks per configuration. 	
	
	Table~\ref{tab:whitebox_ver_conditional_cpr_n1000_allpair_spksim_v3} shows that direct gradient access makes TruMix's output highly 	susceptible to unconstrained target manipulation. PGD-20 increases TruMix's TMR from $75.48\%$ at
	$\epsilon=0.0005$ to $97.08\%$ at $\epsilon=0.004$, while AutoAttack increases it to $99.18\%$ at $\epsilon=0.002$. 	However, the majority of these target successes do not satisfy all
	validity constraints. TruMix's overall VER remains between $22.31\%$
	and $26.75\%$ across all PGD budgets, and $26.44\%$ with  AutoAttack. We can also observe that larger perturbations progressively reduce content and signal quality. These results	indicate that changing the detector's decision is much easier than producing an evasion that simultaneously preserves linguistic content, speaker identity, and signal quality.

As for  Nes2Net-X, at $\epsilon=0.002$, its overall misclassification rate is $61.73\%$ under PGD-20 and $87.03\%$ under AutoAttack, which are lower than TruMix. However, after considering the perceptual constraints,  the difference of the overall VER between the two detectors narrows. 
	
	These results indicate that both systems remain vulnerable to white-box attacks to some extent, and  stronger defense mechanisms need to be explored in the future. 

\begin{table}[!t]
\centering
\caption{White-box adaptive attack.}
\label{tab:whitebox_ver_conditional_cpr_n1000_allpair_spksim_v3}
\footnotesize
\setlength{\tabcolsep}{2.7pt}
\begin{tabular}{@{}llrrrrr@{}}
\toprule
\textbf{Attack} &
\textbf{Budget $\epsilon$} &
\textbf{TMR} &
\textbf{CPR} &
\textbf{SPR} &
\textbf{SNR (dB)} &
\textbf{VER} \\
\midrule
\multicolumn{7}{@{}l}{{\bf TruMix}} \\
PGD-20 &   0.0005 & 
{75.48\%} & {61.36\%} & {99.42\%} &
45.19 & {22.88\%} \\
PGD-20 &  0.0010 & 
{90.30\%} & {59.68\%} & {99.20\%} &
40.25 & {26.75\%} \\
PGD-20 &  0.0020 & 
{95.38\%} & {57.78\%} & {98.87\%} &
34.94 & {26.63\%} \\
PGD-20 &  0.0040 & 
{97.08\%} & {54.99\%} & {98.18\%} &
29.27 & \textbf{22.31\%} \\
AutoAttack &   0.0020 &
{99.18\%} & {56.65\%} & {98.17\%} &
33.27 & \textbf{26.44\%} \\
\midrule
\multicolumn{7}{@{}l}{\bf Nes2Net-X} \\
PGD-20 &   0.0005 & 
{50.10\%} & {61.58\%} & {99.58\%} &
44.01 & {20.63\%} \\
PGD-20 &   0.0010 & 
{59.08\%} & {59.07\%} & {99.29\%} &
39.31 & {20.88\%} \\
PGD-20 & 0.0020 & 
{61.73\%} & {57.03\%} & {98.83\%} &
33.96 & {19.31\%} \\
PGD-20 &  0.0040 & 
{60.58\%} & {53.84\%} & {98.03\%} &
28.26 & \textbf{15.00\%} \\
AutoAttack & 0.0020 & 
{87.03\%} & {52.22\%} & {98.58\%} &
33.45 & \textbf{19.44\%} \\
\bottomrule
\end{tabular}
\end{table}

\section{Conclusion} \label{sec:con}

This work shows that audio authenticity cannot be adequately characterized by a single real-or-synthetic label when speech and acoustic context can be generated or altered independently. To address this limitation, we developed TruMix, a component-level forensic framework that distinguishes eight configurations of speech presence and speech/background authenticity. Its multi-path design examines complementary evidence from speech, background
sound, the complete mixture, and low-level generative artifacts. We also curated MixedTruth, a 4,917-hour benchmark designed to evaluate fine-grained attribution covering diverse generators, content, and mixing conditions. Across extensive evaluations, TruMix achieves high component-level detection accuracy, consistently outperforms existing synthetic-speech detectors in binary detection, and retains strong
performance across four tiers of attack.   Future work will
explore defenses that further strengthen robustness against
adaptive, previously unseen, and more sophisticated attacks.




\cleardoublepage
\appendix
\section*{Ethical Considerations}

This work develops a defensive method for detecting synthetic speech and synthetic background audio. The principal stakeholders include speakers represented in the source corpora, dataset maintainers, users whose recordings may be analyzed, detector developers, and parties relying on audio-authenticity decisions.

\textbf{Data use, privacy, and representativeness.}
MixedTruth is constructed programmatically from previously released speech, anti-spoofing, and environmental-audio corpora obtained under their respective access and research-use terms. We recruited no new participants and collected no new private recordings. Nevertheless, public availability does not imply consent to every downstream synthetic-audio use. We therefore do not redistribute third-party audio when prohibited by its license; such data must be obtained from the original providers. Any released manifests and processing code will retain the corresponding source and licensing information. The benchmark may also inherit demographic, linguistic, and recording-condition biases from its source corpora. Consequently, our results should not be interpreted as uniform across speakers, languages, accents, devices, or acoustic environments.

\textbf{Dual use and safeguards.}
The black-box, gray-box, and white-box attacks used for robustness evaluation could also help adversaries evade synthetic-audio detectors. We evaluate these attacks only against models and recordings under our control, without probing third-party production services or user accounts. We present them to expose component-specific weaknesses and support defensive evaluation, rather than to facilitate impersonation or deceptive-media deployment.

\textbf{Operational use.}
TruMix is not a definitive authenticity oracle. False positives may wrongly cast doubt on genuine recordings, while false negatives may create unwarranted confidence in manipulated audio. Its predictions should therefore be treated as one forensic signal and combined with human review, provenance checks, contextual evidence, and independent corroboration before consequential decisions.

\cleardoublepage
\bibliographystyle{plain}
\bibliography{reference}

\end{document}